\documentclass[twocolumn,article,superscriptaddress,showpacs]{revtex4-1}

\usepackage[utf8]{inputenc}
\usepackage{color}
\usepackage{amsmath}
\usepackage{amssymb}
\usepackage{graphicx}
\usepackage{esint}
\usepackage[unicode=true,
 bookmarks=false,
 breaklinks=true,pdfborder={0 0 0},backref=false,colorlinks=true]
 {hyperref}

\makeatletter
\usepackage{epstopdf}\usepackage{subfigure}
\usepackage{float}
\usepackage{amsfonts}
\usepackage{bm}% bold math
\usepackage{subfigure}
\usepackage{braket}
\usepackage{braket}
\usepackage{physics}
\usepackage{xcolor}

\def\LBO{Li$_{2}$B$_{4}$O$_{7}$}
\def\ZZ{$Z_{2}$}
\def\Z{$Z$}

\def\GT{$\mathcal{GT}$}
\def\T{$\mathcal{T}$}
\def\G{$\mathcal{G}$}

\def\Q{$\mathcal{Q}$}
\def\C{$\mathcal{C}$}
\def\Th{$\tilde{\Theta}$}

\makeatother

\begin{document}

\title{%Universal $Z_2$ bridged by double and quad-helicoid surface states\\
\ZZ{} Dirac points with topologically protected multi-helicoid surface states
}

\author{Tiantian Zhang}
\email{zhang.t.ac@m.titech.ac.jp}
\affiliation{Department of Physics, Tokyo Institute of Technology, Ookayama, Meguro-ku, Tokyo 152-8551, Japan}
\affiliation{Tokodai Institute for Element Strategy, Tokyo Institute of Technology, Nagatsuta, Midori-ku, Yokohama, Kanagawa 226-8503, Japan}

\author{Daisuke Hara}
\affiliation{Department of Physics, Tokyo Institute of Technology, Ookayama, Meguro-ku, Tokyo 152-8551, Japan}

\author{Shuichi Murakami}
\email{murakami@stat.phys.titech.ac.jp}
\affiliation{Department of Physics, Tokyo Institute of Technology, Ookayama, Meguro-ku, Tokyo 152-8551, Japan}
\affiliation{Tokodai Institute for Element Strategy, Tokyo Institute of Technology, Nagatsuta, Midori-ku, Yokohama, Kanagawa 226-8503, Japan}
%

%%---- abstract: less than 600 words 
%%---- the main text is 3750 words, excluding abstract, methods, references and figure legends
%%---- # of refs should be less than 50

\begin{abstract}

%
%In Dirac systems with time-reversal (\T)-glide (\G) symmetry $\tilde{\Theta}^2=$ \GT{}$^2=-1$, multi-helicoid surface states~(MHSSs) appear. However, the gauge-invariant formulation and its bulk-surface correspondence remain unestablished.  
%Here, we establish them by introducing \G-protected \ZZ{} topological invariant $\nu$, which, according to previous works, can be nontrivial only in \T-breaking gapped systems. 
%Unexpectedly, 
%in gapless Dirac systems with $\tilde{\Theta}$, $\nu$ can be nontrivial, offering the global topology of the MHSSs. 
%Moreover, $\nu$ also offers the local topology, showing equivalence with \ZZ{} monopole charge \Q{} associated with Dirac points. 
%\Q{} can be simplified to symmetry-based indicators when two vertical \G{} are preserved, and diagnose filling-enforced band insulators in several cases when \T{} is broken. 
%Material candidate \LBO{} together with a list of space groups preserving MHSSs are also proposed.

In some Dirac systems with time-reversal (\T) and glide (\G) symmetries, multi-helicoid surface states (MHSSs) appear, as discussed in various systems like electronic and photonic ones. However, the topological nature and the conditions for appearance of the MHSSs have not been understood. Here we show that MHSSs result from bulk-surface correspondence for the \ZZ{} monopole charge \Q{}, which \textit{cannot} be defined as a local quantity associated with the Dirac point, unlike the \Z{} monopole charge characterizing Weyl points. The previously known formula of \Q{} turns out to be non-gauge-invariant and thus cannot characterize the MHSSs. This shortcoming of the definition of \Q{} is amended by redefining \Q{} as a global topological invariant in $k$-space. Surprisingly, the newly defined \Q{}, characterizing \GT{} invariant gapless systems, is equal to the \G-protected \ZZ{} topological invariant $\nu$, which is nontrivial only in \T-breaking gapped systems. This global definition of \Q{} automatically guarantees appearance of MHSSs even when the Dirac point split into Weyl points or a nodal ring by lowering the symmetry, as long as the \GT{} symmetry is preserved. \Q{} can be simplified to symmetry-based indicators when two vertical \G{} are preserved, and diagnose filling-enforced topological crystalline insulators in several cases when a \T{}-breaking perturbation is induced. Material candidate \LBO{} together with a list of space groups preserving MHSSs are also proposed.

%% exclude abstract, 2500

\end{abstract}

\maketitle

\section{Introduction}
%The bulk-surface correspondence associated with helical surface states are topologically protected by the monopole charge \C{} in Weyl semimetals, while it is not guaranteed in Dirac semimetals due to \C{}=0. 
%However, we show a strict bulk-surface correspondence associated with anti-parallel helical surface states can be obtained for Dirac semimetals under the protection of a new monopole charge \Q{}. Furthermore, the double Fermi arcs for the Dirac points are protected by the global topology of \Q{}, which is originally from the glide-mirror-protected topological invariant. 
%A material with double Fermi arcs is proposed to demonstrate the experimental fingerprint of nontrivial topology for Dirac semimetals and deepen the understanding of topological quantum materials in general.
%

%---------------------------------------------introduction-----------------------
%\section*{Introduction}
Starting from the first proposal of the 
\T-protected topological insulator in 2005~\cite{kane2005z,bernevig2006quantum,fu2007topological}, a vast number of topological (crystalline) insulators~(TI/TCI) were discovered in gapped band structures with different symmetries afterwards~\cite{schnyder2008classification,kitaev2009periodic,chiu2016classification,po2017symmetry,bradlyn2017TQC,song2018quantitative,song2018diagnosis,khalaf2018symmetry}. %, such as \T{}, translation, (glide) mirror and (screw) rotation. 
%In fact, all the topological phases enumerated by their topological invariants can be uniquely mapped from eigenspace to homotopy groups by discrete symmetries, including on-site symmetries~\cite{schnyder2008classification,kitaev2009periodic,chiu2016classification} and crystalline symmetries~\cite{po2017symmetry,song2018quantitative,song2018diagnosis,khalaf2018symmetry}. 
Topological equivalence of two systems can be diagnosed by either the type or the value of topological invariants, associated with diverse topological surface states due to bulk-surface correspondence~(BSC). 
%they do not change the value unless a discontinuous deformation of the space happens. 
For example, TI associated with surface Dirac cone(s) are proposed with topological invariant $z_2$=1 in Ref.~\cite{fu2007topological}. 
%a $Z_2$ type topological invariant $\nu$ is proposed in \T-breaking gapped systems with \G{}- symmetry~\cite{fang2015new, shiozaki2015z}, where the topological crystalline insulator~(TCI) phase can be obtained when $\nu=1$, associated with a single unpinned surface Dirac cone located along \G-invariant lines. 
%Particularly, for \T-breaking spinless systems with two vertical \G, another $Z_2$ type topological invariant, i.e., symmetry-based indicator $\mu_2$, is offered to diagnose the \G-protected TCI phase~\cite{ono2018unified}. %which was conjectured to be related with the filling-enforced band insulators in some cases.  
As for topological semimetals, their band degeneracies are protected by various topological invariants associated with disparate BSC. 
Such as the Berry phase $\gamma$ for the nodal line/ring with drumhead surface states~\cite{burkov2011topological,fang2016topological,bian2016drumhead,chan20163,deng2019nodal}, the $Z$ type monopole charge $\mathcal{C}$ for Weyl points with helical surface states~\cite{weng2015weyl,Weyl_Taas,Weyl_experiments,Weyl_exp,fang2016hss,yang2018ideal,chen2016photonic,he2020observation,zhang2018double,zhang2020twofold} and \ZZ{} type monopole charge \Q{} for Dirac points with MHSSs such as double (quad-)helicoid surface states (DHSSs/QHSSs)~\cite{yang2014classification,morimoto2014weyl,shiozaki2015z,fang2016hss,gorbar2015surface,cai2020symmetry,cheng2020discovering}. 
In particular, DHSSs/QHSSs associated with Dirac points realized in \GT{}-preserving systems are particularly interesting. Because they originate from Dirac points, which are composites of a pairs of Weyl points with \C=$\pm1$, they can escape from gap opening only under some conditions. 
Nevertheless, these conditions have not been identified because their topological nature has not been understood so far. 
Therefore, it is not known how to realize DHSSs/QHSSs in general systems, such as electronic and photonic systems. 
%Since there is no gauge-invariant definition for \Q{}, the lack of strict proof for its BSC is also not a surprise. 
Furthermore, two more topological invariants $\nu$~\cite{fang2015new, shiozaki2015z} and $\mu_2$~\cite{ono2018unified} can be defined in terms of \G{} when the system is fully gapped, but there is no research studying their relationships with \Q{} so far. 

In this paper, we point out that the previous definition of the \ZZ{} monopole charge \Q{} is not gauge invariant, and show a new gauge-invariant definition for \Q{}. Thereby, we can show the BSC between nontrivial value of \Q{} and MHSSs. 
In Section~\ref{sectionII}, we start the discussion for conventional Dirac points composed of two Weyl points with opposite $\mathcal{C}$ in Dirac semimetal systems, in which no gapless helical surface states are guaranteed, and then discuss \ZZ{} Dirac points carrying monopole charge \Q{} protected by $\tilde{\Theta}$ in Section~\ref{sectionIIIA} and Section~\ref{sectionIVB}. Here, we give an amended gauge-invariant definition of \Q{}. 
We also show that the \ZZ{} monopole charge \Q{} associated with Dirac points under \G{} and \T{} symmetry cannot be defined as a local quantity, as opposed to previous works, and, surprisingly, it is equal to $\nu$, which characterizes a \G-protected TCI phase without \T-symmetry in Section~\ref{sectionIIIB}. 
With this newly defined gauge-invariant \Q{} (=$\nu$), we establish BSC for Dirac points with both DHSSs and QHSSs for \GT-preserving systems in Section~\ref{sectionIIID} and Section~\ref{sectionIVC}. 
This global definition of \Q{} automatically guarantees appearance of MHSSs even when the Dirac point split into Weyl points or a nodal ring by lowering the symmetry, as long as the \GT{} symmetry is preserved. 
We find that QHSSs can be only retained in spinless systems, and it vanishes in spinful systems due to the ill-defined \Q{} in Section~\ref{sectionIV}. 
The first \ZZ{} Dirac material candidate \LBO{} and a list of space groups with QHSSs will be offered in Section~\ref{sectionV}. 
Therefore, our theory has established conditions to guarantee DHSSs/QHSSs, which can lead to their realization in a broad range of physical systems.

\begin{center}
\begin{figure*}[ht]
\includegraphics[scale=0.82]{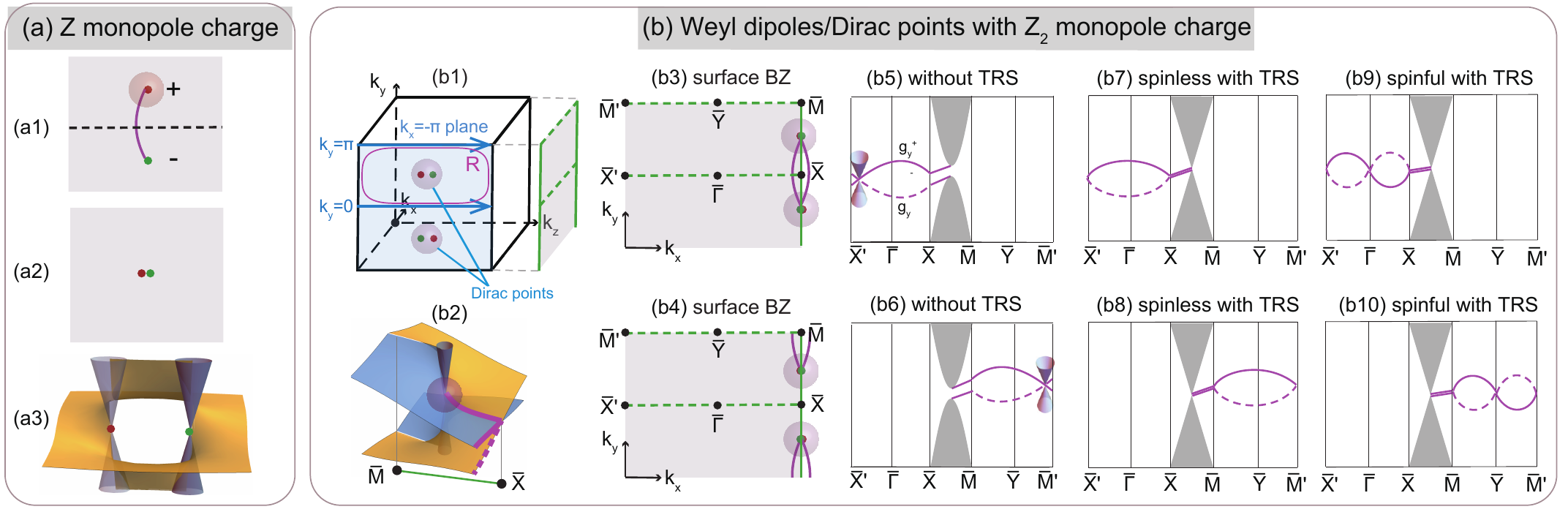}\caption{\ZZ{} Dirac points under \G- and \T-symmetries and associated DHSSs. 
%\textbf{Fermi arcs and surface-state connections for Weyl points and Dirac points.} 
(a1) Fermi arcs for Weyl points with opposite monopole charges $\mathcal{C}$, which can be defined either by the sphere enclosed the Weyl point or by a 2-dimensional plane marked by the dashed line. 
(a2) Fermi arcs do not necessarily exist for a conventional Dirac point composed of a pair of Weyl points with opposite $\mathcal{C}$. 
(a3) Helical surface states for a pair of Weyl points carrying opposite $\mathcal{C}$.  
(b1) Locations for \ZZ{} Dirac points (Weyl dipoles) on the $k_x=-\pi$ plane (blue plane) satisfying \Th$_y^2=-1$ in the bulk BZ. 
(b2) Double-helicoid surface states contributed by \ZZ{} Dirac points, which will be projected along $\bar{M}$-$\bar{X}$ on the surface BZ. The gray cone is the bulk \ZZ{} Dirac band. Blue and yellow sheets are the anticrossing helical surface states. Purple lines show the surface-state connections along $\bar{\Gamma}$-$\bar{X}$-$\bar{M}$ directions.  
(b3) Fermi arcs for \ZZ{} Dirac points (Weyl dipoles) on the \G$_y$-preserved surface BZ, which corresponds to the gray plane shown in (b1). The Fermi arcs will change to (b4) when the energy changes. 
(b5-b6) Two possible surface-state connections with $\nu_y=1$ (=$\nu_y^{\mathrm{surface}}$) in the \T-breaking system defined by single \G$_y$, with the single surface Dirac cone located along different \G$_y$-invariant lines. 
``TRS'' represents for time-reversal symmetry.
(b7-b8) Two possible topological surface-state connections for \ZZ{} Dirac points in the spinless and \T-preserving case. 
(b9-b10) Two possible topological surface-state connections for \ZZ{} Dirac points in the spinful and \T-preserving case. 
Surface states shown in (b7-b10) are all in the double-helicoid shape shown in (b2). 
\textbf{}
\label{fig:ZZ}}
\end{figure*}
\end{center}

%%%----------------  charge-Z/Z2-----------%%%
\section{Monopole charges \C{} for Weyl and Dirac points}
\label{sectionII}
Monopole charge $\mathcal{C}$ can be defined by the Chern number on a sphere enclosing the Weyl point, or on a two-dimensional (2D) plane in the 3D Brillouin zone (BZ) marked by the dashed line in Fig.~\ref{fig:ZZ} (a1). 
The former definition shows the local property of $\mathcal{C}$, while the latter one offers the global topology to understand the influence of $\mathcal{C}$ to the whole BZ, giving rise to gapless surface states connecting two Weyl points with opposite \C{}, as shown in Fig.~\ref{fig:ZZ} (a3). 
Fermi arcs are isoenergetic surface states, and they can disappear when those two Weyl points are projected onto the same momentum on the surface BZ or when they are forced to coalesce into a Dirac point by additional symmetries~\cite{wang2012dirac,liu2014discovery,neupane2014observation,liu2014stable,borisenko2014experimental,steinberg2014bulk,zhang2015breakdown,xu2015observation,wang2016three,tang2016dirac,kargarian2016surface,guo2017three,li2017topological,slobozhanyuk2017three,guo2017three,yao2018topological,kargarian2018deformation,bao2018discovery,guo2019observation,le2018dirac,wu2019fragility,zhang2019multiple,yang2019realization}, as shown by the purple solid line in Figs.~\ref{fig:ZZ} (a1-a2). 

Although the monopole charge $\mathcal{C}$ will vanish for a Dirac point~\cite{young2012dirac,wang2012dirac,Top_CDAS,yang2014classification,mullen2015line}, a new \ZZ{} topological invariant \Q{} can be defined when an anti-unitary operator $\tilde{\Theta}$=\GT{} with $\tilde{\Theta}^2=-1$ is present~\cite{fang2016hss}, and its corresponding topological bulk degeneracies are named \ZZ{} Weyl dipoles or \ZZ{} Dirac points (Detailed discussions on Weyl dipoles are in \cite{supp}). 
\ZZ{} Dirac points always appear in pairs and are located at \T-related momenta in the BZ. 
Since \ZZ{} Weyl dipoles and \ZZ{} Dirac points can be transformed mutually by  $\tilde{\Theta}$-preserved perturbations, we will use \ZZ{} Dirac points as an example to show the global topology of \Q{} in the following. 
%We also note that \ZZ{} nodal rings protected by \Th{} are not discussed in this work, since they preserve different type of surface states.

\section{{monopole charge \Q{} in systems with single \GT{}}}
\label{sectionIII}
In this section, we will show that the definition of $Z_2$ monopole charge \Q{} given in Ref.~\cite{fang2016hss} in systems with single \G{} and \T{} actually depends on a gauge, and will redefine \Q{} in a gauge-independent way. 
This eventually shows that \Q{} cannot be defined locally in $k$-space, but that \Q{} shows global topology in $k$-space.

\subsection{Redefinition of monopole charge \Q}
\label{sectionIIIA}

%As a result of the bulk-surface correspondence, surface states are always associated with topological states of various types. To date, those three topological invariants have not been well understood yet for the systems with two glide mirrors, and also for the and the bulk-surface correspondence of those systems, which was a frequently stated problems in several studies. 
%, charge-\ZZ{} can be also defined on the 2D plane parallel to the glide plane~\cite{fang2016hss}, and changes its value when the 2D plane cross the Dirac points. 

\begin{figure}[h]
\centering
\includegraphics[scale=0.45]{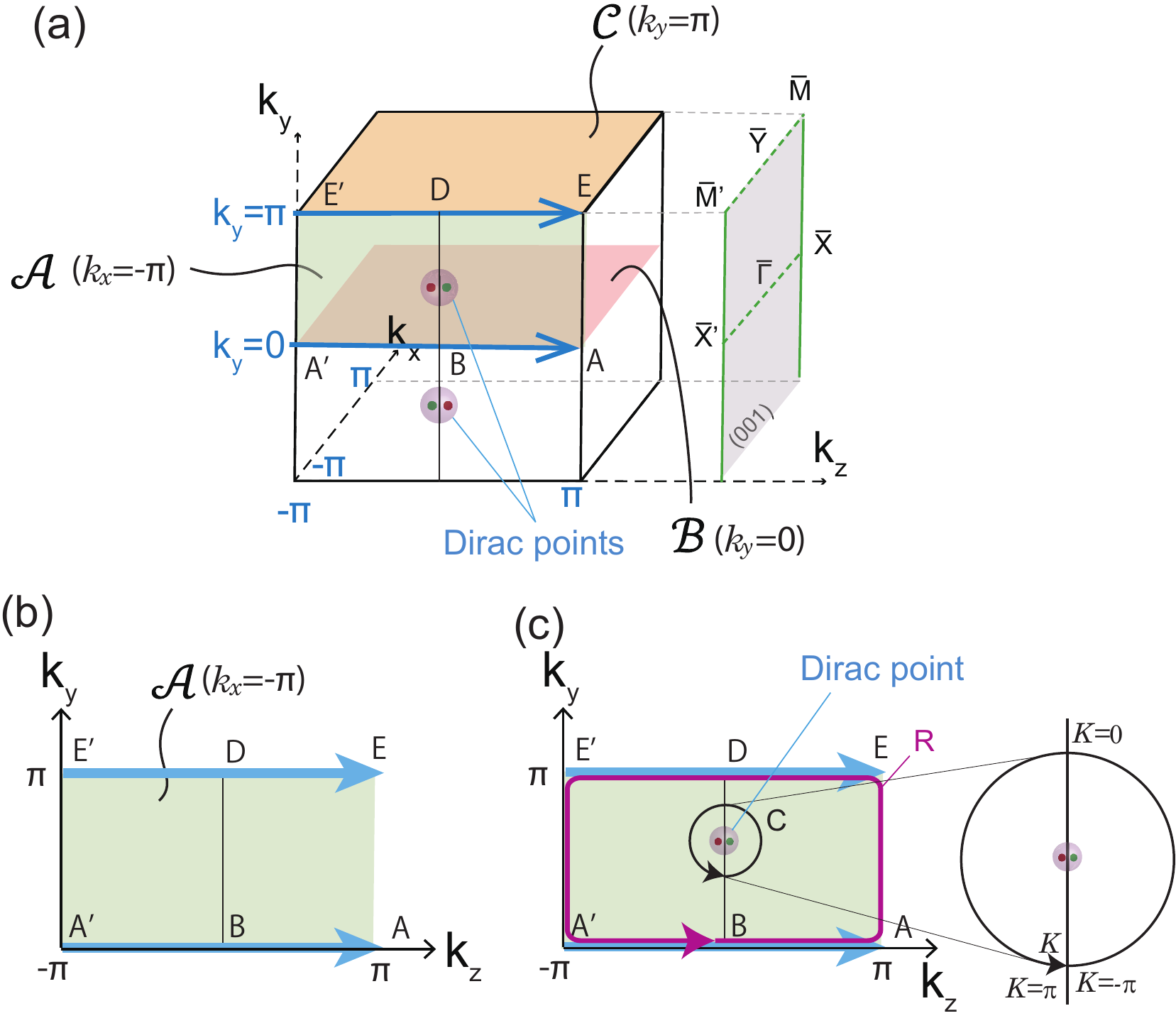}
\caption{
%Glide-protected topological invariant $\nu$ and $Z_2$ monopole charge \Q{} for a system with single glide symmetry. 
(a) Brillouin zone for
the space group \#7, which only has a glide symmetry, and the colored planes are used for the calculation of $\nu$. 
Projection onto the (001) plane is also shown. 
The region $\mathcal{A}$, $\mathcal{B}$ and $\mathcal{C}$ are defined by $\mathcal{A}$=\{$\mathbf{k}|k_x=-\pi, 0\leq k_y\leq \pi, -\pi\leq k_z\leq \pi$\}, 
$\mathcal{B}$=\{$\mathbf{k}|k_x= 0, -\pi\leq k_y\leq \pi, -\pi\leq k_z\leq \pi$\}, 
$\mathcal{C}$=\{$\mathbf{k}|k_x=\pi, -\pi\leq k_y\leq \pi, -\pi\leq k_z\leq \pi$\}. 
(b) Slice of the Brillouin zone at $k_x=-\pi$. If the system
has both time-reversal and glide symmetries, the \ZZ{} invariant 
$\nu$ is shown to be equal to the difference of $\tilde{\Theta}$-polarization, calculated as a Berry phase along the two blue lines. 
(c) When the system has a Dirac point either on the 
DB line or on the EA line, and is gapped in the rest of the region $\mathcal{A}$, the $Z_2$ monopole charge \Q{} for the
Dirac point can be well defined. 
In the previous work, \Q{} is defined along the circle C parameterized. 
To guarantee the gauge invariance of \Q, we 
need to deform the circle C to the rectangle R (purple line), which makes \Q{} to be equal to $\nu$.}
\label{fig:bz7}
\end{figure}

We consider systems with one glide symmetry $\mathcal{G}_{y}=\{M_y|\frac{1}{2}00\}$, where $M_y$ represents the mirror reflection with respect to the $xz$ plane. 
Systems with only a $\mathcal{G}_{y}$ symmetry correspond to \#7, and their BZs are shown in both Fig.~\ref{fig:ZZ} (b1) and Fig.~\ref{fig:bz7} (a).  
Let $\tilde{\Theta}_y$ denote an antiunitary operator $\tilde{\Theta}_y$=\T\G$_{y}$, which  leads to  $\tilde{\Theta}_{y}^2=e^{-ik_x}$. Therefore 
$k_x=+\pi/-\pi$ is a special plane where $\tilde{\Theta}_{y}^2=-1$ is 
satisfied, as shown by the blue plane in Fig.~\ref{fig:ZZ} (b1) and green plane (region $\mathcal{A}$) in Figs.~\ref{fig:bz7} (a-c), 
and here we consider the case with 
two \ZZ{} Dirac points on the $k_x=-\pi$ plane in the 3D BZ. They are related by \T{} symmetry, and 
%\textcolor{gray}{When the system only have two pairs of Weyl dipoles, those four Weyl points can locate at any momenta with the only restriction to meet the $\tilde{\Theta}_y$ symmetry, as shown in Fig.~\ref{fig:ZZ} (b1). However, if additional symmetries like \P{} or screw rotation are present, two Weyl points inside each Weyl dipole will move to each other and meet together at the $\tilde{\Theta}_y^2=-1$ line forming two \ZZ{} Dirac points, as shown in Figs.~\ref{fig:ZZ} (b1-b2).} 
so they cannot appear at time-reversal-invariant momenta (TRIM). Thus, \ZZ{} Dirac points are usually located on high-symmetry lines for systems only has one \G.

Here, we consider a Dirac point on the $\tilde{\Theta}$-invariant line with $\tilde{\Theta}^2=-1$, i.e., the E-A or D-B line on the $k_x=-\pi$ plane shown in Fig.~\ref{fig:bz7} (c). 
A $Z_2$ monopole charge \Q{} is associated with the Dirac point,
defined in terms of wavefunctions on a $\tilde{\Theta}$-symmetric sphere enclosing the Dirac point, according to Ref.~\cite{fang2016hss}.
The formulation of \Q{} closely follows that of the $Z_2$ topological invariant for time-reversal invariant 
topological insulators without inversion symmetry \cite{Fu2006prb74}, but with a replacement of \T{} by $\tilde{\Theta}$:
\begin{align}
& (-1)^{\mathcal{Q}}=\frac{\mathrm{Pf}[W(0)]}{\sqrt{\mathrm{Det}[W(0)]}}\frac{\mathrm{Pf}[W(\pi)]}{\sqrt{\mathrm{Det}[W(\pi)]}}
\label{eq:QWW}
\end{align}
where the matrix $W_{mn}(K)$ is defined by
\begin{align}
& W_{mn}(K)=\langle u_m(-K)|\tilde{\Theta}|u_n(K)\rangle.
\end{align}
with $m$ and $n$ running over the occupied bands \cite{fang2016hss}.
Here we have introduced a coordinate $K$ ($-\pi\leq K\leq \pi$) along the circle C, as shown in Fig.~\ref{fig:bz7}(c), so that $\tilde{\Theta}$ transforms $K$ to 
$-K$.
It was claimed that \Q{} $\pmod{2}$ is gauge invariant in Ref.~\cite{fang2016hss}; nevertheless, we point out that this discussion of gauge invariance is not correct.
As opposed to the discussion in the $Methods$ of Ref.~\cite{fang2016hss}, some gauge transformation (e.g. multiplying a phase factor $e^{iK}$ to one eigenstate) changes the branch choice of the square root in Eq.~(\ref{eq:QWW}) and will alter the value of \Q{} by unity. Thus, we hereby reexamine the definition of the $Z_2$ monopole charge \Q.

%%%-------

Similar to the \T-polarization \cite{Fu2006prb74}, we here note that the crucial condition for \Q{} is $\tilde{\Theta}^2=-1$, which is satisfied on the plane $k_x=-\pi$. 
Thus, among the wavefunctions on the sphere surrounding the Dirac point, only those on the circle C lying on $k_x=-\pi$ plane are relevant for the definition in Eq.~(\ref{eq:QWW}). 
Since there is a gauge transformation for wavefunctions along circle C, which will change \Q{} by unity (see the last paragraph of Sec.~IIIB in Ref.~\cite{Fu2006prb74}), this definition is not sufficient to make the $Z_2$ monopole charge \Q{} to be well defined. 

To make the value of \Q{} gauge independent modulo 2, we need to make some constraints on the gauge choice. It is achieved by imposing that the gauge of the wavefunctions is continuous within the region $\mathcal{A}$. It allows us to enlarge the circle C to a rectangle R in a $\tilde{\Theta}$-symmetric way in the definition of \Q{} (see Fig.~\ref{fig:ZZ} (b1) and Fig.~\ref{fig:bz7}(c)). 
It is possible only when the bulk band structure on the $k_x=-\pi$ is gapped except for the Dirac points considered. 
Since wavefunctions should be periodic along the $k_z$ direction, i.e., equal between the
two edges of the rectangle, EA and E$^{\prime}$A$^{\prime}$, the above gauge transformation altering \Q{} by unity is now prohibited. 
Under this gauge condition, similarly to the argument for inversion-asymmetric \ZZ{} TIs in Ref.~\cite{Fu2006prb74}
, the $Z_2$ monopole charge is rewritten to be a difference in ``$\tilde{\Theta}$-polarization'' between $k_y=0$ and $k_y=\pi$, and we conclude
\begin{align}
\mathcal{Q}&= P_{\tilde{\Theta}}(k_y=\pi)-P_{\tilde{\Theta}}(k_y=0)  \pmod{2},
\end{align}
where $P_{\tilde{\Theta}}(k_y)=\frac{1}{2\pi}(\gamma^+_{\mathrm{L}}-\gamma^-_{\mathrm{L}})$.  $\gamma_{\mathrm{L}}$ is the Berry phase with the integral path $\mathrm{L}$ taken as $k_y=0$ and $k_y=\pi$ lines on the $k_x=-\pi$ plane and ``$\pm$'' represent the sectors with positive or negative glide eigenvalues for the Bloch wavefunctions. We call $P_{\tilde{\Theta}}(k_y)$ $\tilde{\Theta}$-polarization because \Th{}$_y$ switches those two glide sectors, in an analogy to the time-reversal polarization in Ref.~\cite{Fu2006prb74}. Thus, as opposed to previous works, \Q{} can be \textit{only} defined globally in $k$-space in Eq.~(\ref{eq:nugamma}). 
It is in strong contrast with the locally defined monopole charge \C{} for Weyl points. 

To summarize, we found that the definition of the $Z_2$ monopole charge \Q{} for the Dirac point in a previous work in terms of wavefunctions on a circle C surrounding the Dirac point is gauge dependent and ill defined. 
To make \Q{} well defined, we need to impose a condition that the system is gapped everywhere in the region $\mathcal{A}$ ($k_x=-\pi$, $-\pi\leq k_z\leq \pi$, $0\leq k_y\leq \pi$), except for the
Dirac point. Under this condition, the \ZZ{} monopole charge is now well defined in terms of wavefunctions along the rectangle R. 
In this sense, the $Z_2$ monopole charge cannot be defined locally but has a global nature. 
We also note that a simplified formula for \Q{} in systems with an additional twofold rotation or screw symmetry is given in terms of eigenvalues of those rotation symmetries in Ref.~\cite{fang2016hss}, but it is valid only when the system satisfies the conditions discussed above.

All the discussions above are in spinless systems, and we note that the definition of \Q{} remains the same in spinful systems because $\tilde{\Theta}^2=e^{-ik_x}$ remains true also in spinful systems with single glide symmetry. 
We further notice that \Q{} is equal to the global \G-protected \ZZ{} topological invariant $\nu$ mathematically and physically in both the spinless and spinful systems, which will be discussed in detail in the next section.

%%%
%%%
%%%% add for PRR 

\subsection{Topological invariant $\nu$ for gapped systems with \G{}}
\label{sectionIIIB}

In spinless systems with glide symmetry $\mathcal{G}_y=\{M_y|\frac{1}{2}00)\}$, %($\mathcal{G}_y$ is defined by $\{M_y|\frac{1}{2}00\}$ in the main text, but shares the same results with the one defined in the supplementary)
 eigenstates on glide-invariant planes $k_y = 0$ and $k_y = \pi$ are classified into two glide sectors with
\begin{equation}
g_\pm (k_x) = \pm e^{- ik_x /2}.
\label{eq:glide_eigenval}
\end{equation}
For three-dimensional (3D) spinless systems with a gapped band structure, a $Z_2$ type topological invariant \cite{fang2015new,shiozaki2015z,Kim2019prb100} can be defined by the glide symmetry as
\begin{align}
\nu &= \frac{1}{2\pi} \left[ \int_{\mathcal A} F_{yz} dk_y dk_z + \int_{\mathcal{B} } F^-_{zx} dk_z dk_x \right. \nonumber \\
 &\ \ \ \ \ - \left. \int_{ \mathcal{C}} F^-_{zx} dk_z dk_x \right] \nonumber \\
& \ \ \ - \frac{1}{\pi} \left( \gamma^+_{\mathrm{A}^\prime\mathrm{BA}} + \gamma^+_{\mathrm{ED}\mathrm{E}^\prime} \right) \pmod{2},
\label{eq:nu}
\end{align}
where $\mathcal{A}$, $\mathcal{B}$ and $\mathcal{C}$ are the integral regions shown in Fig.~\ref{fig:bz7}(a). 
In this formula, we have defined the Berry connections
\begin{align}
& \bm{A}(\bm k) \equiv \sum_{n\in \mathrm{occ}} i \bra{u_{n\bm k}} \nabla_{\bm k} \ket{u_{n\bm k}} , \\
& \bm{A}^\pm(\bm k) \equiv \sum_{n\in \mathrm{occ}} i \bra{u^\pm_{n\bm k}} \nabla_{\bm k} \ket{u^\pm_{n\bm k}} ,
\end{align}
and the corresponding Berry curvatures 
\begin{align}
F_{ij} (\bm{k}) = \partial_{k_i} A_j (\bm{k}) - \partial_{k_j} A_i (\bm{k}), \\
F^\pm_{ij} (\bm{k}) = \partial_{k_i} A^\pm_j (\bm{k}) - \partial_{k_j} A^\pm_i (\bm{k}) ,
\end{align}
where the summation $\sum_{n\in \mathrm{occ}}$ is over the occupied states with band index $n$, and $\ket{u^\pm_{n\bm k}}$ are the Bloch wavefunctions within the positive/negative glide sectors of $g_\pm (k_x) = \pm e^{-ik_x/2}$.
The Berry phase $\gamma^{\pm} (\bm{k})$ is defined along a closed path $\lambda$ as
\begin{equation}
\gamma^{\pm}_\lambda (\bm{k}) = \oint_\lambda \bm{A}^{\pm} (\bm{k}) \cdot d\bm{k}.
\end{equation}
In Eq.~(\ref{eq:nu}),  $\gamma^+_{\mathrm{A}^\prime\mathrm{BA}}$ and $\gamma^+_{\mathrm{E}\mathrm{DE}^\prime}$ are Berry phases with  the paths $\lambda$ 
taken as straight lines with $\mathrm{A}'\rightarrow\mathrm{B}\rightarrow\mathrm{A}$ and $\mathrm{E}\rightarrow\mathrm{D}\rightarrow\mathrm{E}'$, respectively, 
where the corresponding high-symmetry points are shown in Fig.~\ref{fig:bz7}. We note that the wavefunctions are periodic along the $k_z$ direction. 
It is noted that for convenience, we take the glide operation to be $\mathcal{G}_y=\{M_y|\frac{1}{2}00\}$, which is different from $\{M_y|00\frac{1}{2}\}$ taken in the previous papers \cite{fang2015new, shiozaki2015z,Kim2019prb100}. 

%%%
We now consider systems with time-reversal symmetry (\T). 
In this case, integrals on the $\mathcal{B}$ and $\mathcal{C}$ surface will vanish, because 
the Berry curvature will change its sign, while the glide sector will be unchanged under \T{}. The integral on $\mathcal{A}$ is rewritten as:
\begin{align}
\int_{\mathcal A} F_{yz} dk_y dk_z = \gamma_{\mathrm{A}^\prime\mathrm{BA}} + \gamma_{\mathrm{ED}\mathrm{E}^\prime} ,
\end{align}
where the gauge of the wavefunctions is taken to be continuous between $k_z=-\pi$ and $k_z=\pi$.
Thus, Eq.(\ref{eq:nu}) can be written as:
\begin{align}
\nu&= \frac{1}{2\pi} \left( \gamma^-_{\mathrm{A}^\prime\mathrm{BA}} -\gamma^+_{\mathrm{A}^\prime\mathrm{BA}}
+ \gamma^-_{\mathrm{ED}\mathrm{E}^\prime} -\gamma^+_{\mathrm{ED}\mathrm{E}^\prime} \right) \nonumber\\
&=P_{\tilde{\Theta}}(k_y=\pi)-P_{\tilde{\Theta}}(k_y=0)\pmod{2},
\label{eq:nugamma}
\end{align}
where we define
\begin{align}
P_{\tilde{\Theta}}(k_y)=\frac{1}{2\pi}(\gamma^+_{\mathrm{L}}-\gamma^-_{\mathrm{L}})\ 
 \  (k_y=0,\pi),
\label{eq:Pthetapm}
\end{align}
with $\mathrm{L}=(\mathrm{A}^\prime\mathrm{BA},\mathrm{E}^\prime\mathrm{DE})$
 being a straight path along the $k_x$ direction with fixed $k_y(=0,\pi)$ on the $k_x=-\pi$ plane. 
It is a polarization difference between two glide sectors along the 1D subspaces marked by the blue arrows in Fig.~\ref{fig:bz7}(b). 
The antiunitary symmetry $\tilde{\Theta}\equiv G_y$\T{} gives rise to double degeneracy on the DB and EA lines due to \Th{}$^{2}=-1$. Then the glide sectors on the $k_x=-\pi$ plane are switched by $\tilde{\Theta}$.

Next, we show that the definition of Eq.~(\ref{eq:Pthetapm}) can be extended to arbitrary values of $k_y$ ($0<k_y<\pi$). By classifying eigenstates of occupied states
into two sets I and II, where the two sets are mutually tranformed by $\tilde{\Theta}$, 
we define
\begin{align}
P_{\tilde{\Theta}}(k_y)=\frac{1}{2\pi}(\gamma^{\mathrm{I}}_{\mathrm{L}}-\gamma^{\mathrm{II}}_{\mathrm{L}})\ 
 \  (0\leq k_y\leq\pi).
\label{eq:Pthetapm2}
\end{align}
This is reduced to Eq.~(\ref{eq:Pthetapm}) for
$k_y=0,\pi$. 
We note that $\tilde{\Theta}$-polarization is an integer, and so is $\nu$. 
Since \T-polarization is gauge dependent~\cite{Fu2006prb74}, $\tilde{\Theta}$-polarization will also be. 
However, the difference of $\tilde{\Theta}$-polarization between 
$k_y=0$ and $k_y=\pi$ is defined in terms of modulo 2, which is gauge independent due to the continuous gauge choice on the $\mathcal{A}$ plane. 

Therefore, when the system is fully gapped and \T-symmetric, $P_{\tilde{\Theta}}(k_y)$ is a quantized integer and a continuous function of $k_y$. Thus, from Eq.~(\ref{eq:nugamma}), $\nu$ vanishes in gapped systems with \T{} as discussed in Refs.~\cite{fang2015new, shiozaki2015z}. 
In other words, the \G-protected TCI phase requires breaking of \T-symmetry. 

\subsection{Topological invariant $\nu$ for Dirac systems with \G{} and \T}
\label{sectionIIIC}
Although this \G-protected topological invariant $\nu$ is originally defined for fully gapped systems, 
we propose that it can be extended to gapless systems with \T by Eq.~(\ref{eq:nugamma}), 
as long as the system is gapped along  the two paths $\mathrm{A}^\prime\mathrm{BA}$ and $\mathrm{ED}\mathrm{E}^\prime$ in Eq.~(\ref{eq:nugamma}).
For example, by adding a twofold rotation along the $y$ axis, doubly degenerate states can cross along the E-A line or D-B line and the system become gapless, which is the case considered in 
Ref.~\cite{fang2016hss}. In this case, $\nu$ can be nontrivial even when \T{} is preserved, and \Q{} is equal to the global \G-protected \ZZ{} topological invariant $\nu$ mathematically and physically in both the spinless and spinful systems:

\begin{align}
\mathcal{Q}&=\nu.
\end{align}

This is counterintuitive since $\nu$ is originally defined for fully gapped systems and it vanishes when the system is \T-invariant \cite{fang2015new, shiozaki2015z}, while \Q{} is associated with the Dirac point in \T-preserving gapless systems. 
We propose in this paper that $\nu$ is also well-defined by Eq.~(\ref{eq:nugamma}), even for gapless systems, as long as the system is fully gapped along the two blue lines marked in Fig.~\ref{fig:ZZ} (b1). It is nontrivial when the system is \T-invariant and have Dirac cones on the $\bar{M}$-$\bar{X}$ line. 
The nontrivial \Q{} ($=\nu$) results in topological surface states on the $\mathcal{G}_y$-preserving (001) surface, which will be discussed in the next subsection.

%%%%% add for PRR
\subsection{{BSC for \Q{} with single \G{}}} 
\label{sectionIIID}

Observing topological states on the surface is the simplest and most straightforward way to demonstrate the topology of the bulk states, due to the BSC. 
Here we establish BSC for
a nonzero \Q{} with single \G{} in both the spinless and spinful systems. In the present case, 
two Dirac points are projected onto the 
$\bar{M}$-$\bar{X}$ line satisfying $\tilde{\Theta}_y^2=-1$. The Fermi energy is set at the Dirac point. 
In this subsection, we will show that 
Fermi arcs for a nonzero \Q{} have two possibilities shown in Figs.~\ref{fig:ZZ} (b3-b4) on the \G$_y$-preserved surface BZ, where 
the surface states extend either toward $\bar{M}$ or $\bar{X}$. 
The green dashed lines are glide-invariant ones, and the green solid lines are $\tilde{\Theta}_y$-invariant ones, with
$\tilde{\Theta}_y^2=-1$, giving rise to double degeneracy for surface states. 

In gapped systems with \G{}$_y$ symmetry, the bulk \ZZ{} topological invariant
$\nu_y$ is well-defined, and by the BSC, it is equal to 
the surface \ZZ{} topological invariant $\nu_y^{\mathrm{surface}}$ which characterizes
how the surface states cross the Fermi energy (see {Sec.~S1} for details~\cite{supp}). 
However, in the present gapless system, $\nu_y$ remains well-defined by Eq.~(\ref{eq:nugamma}), while $\nu_y^{\mathrm{surface}}$ becomes ill-defined due to
the bulk gap closing along $\bar{M}$-$\bar{X}$. 
To establish BSC, in the present gapless system, 
firstly, we
slightly break the \T{} symmetry to open a small bulk gap at the Dirac point. 
Figures~\ref{fig:ZZ} (b5-b6) are two possible nontrivial surface-state connections with $\nu_y^{\mathrm{surface}}$=1 in gapped band structures without \T~\cite{fang2015new, shiozaki2015z}, which has a single unpinned surface Dirac cone along the glide-invariant lines, $\bar{M}$-$\bar{Y}$-$\bar{M}'$ or $\bar{X}$-$\bar{\Gamma}$-$\bar{X}'$, but not both. 
%As long as the perturbation is nonzero, the surface-state connection remains nontrivial. 
Next, we make the \T-breaking perturbation to be zero; all the states along $\bar{X}$-$\bar{M}$ then becomes doubly degenerate due to \Th$^2=-1$, resulting in two kinds of possible surface-state connections, both without SOC (Figs.~\ref{fig:ZZ} (b7-b8)) and 
with SOC (Figs.~\ref{fig:ZZ} (b9-b10)). 
The corresponding Fermi arcs are shown 
in Figs.~\ref{fig:ZZ} (b3-b4).
Remarkably, the doubly degenerate surface states
should start exactly at the Dirac point, belonging to the DHSSs shown in Fig.~\ref{fig:ZZ} (b2). These DHSSs can be interpreted as
a superposition of surface states from two Weyl points with opposite $\mathcal{C}$, and their intersections are protected along $\bar{X}$-$\bar{M}$ by $\tilde{\Theta}_y^2=-1$, in both the spinless and spinful systems.
Such DHSSs also have a \ZZ{} nature, 
which directly follows from the \ZZ{} nature of $\nu$ in gapped systems \cite{fang2015new, shiozaki2015z}. 
For example, when there are two Dirac points within $0<k_y<\pi$ on the $k_x=-\pi$ plane, two sets of DHSSs 
are expected, but they can be annihilated by a continuous change
of surface states without changing the bulk bands.

\section{{\Q{} with two vertical \G{} in spinless systems}}
\label{sectionIV}

\ZZ{} Dirac systems with two $\tilde{\Theta}$ have more unresolved mysteries, and we find that \Q{} is well-defined \textit{only }in the spinless systems, which is beyond people's expectation~\cite{fang2016hss}. 
In this section, we will show that when additional crystalline symmetries like an additional vertical glide symmetry are present, \Q{} ($=\nu$) can be in a form of a symmetry-based indicator calculated in terms 
of irreducible representations at high-symmetry points. 

For example, systems with \#110 have two glide symmetries perpendicular with each other, i.e., $\mathcal{G}_x=\{M_x|0\frac{1}{2}0\}$ and $\mathcal{G}_y=\{M_y|00\frac{1}{2}\}$. Those two vertical glide symmetries lead to \Th$_x^2=-1$ at $k_y=\pi$ and \Th$_y^2=-1$ at $k_z=\pi$, resulting in \ZZ{} Dirac points at non-TRIM high-symmetry points, e.g. $P$ and $P^{\prime}$.  
The original formula for the \ZZ{} glide invariant $\nu$ is expressed as a sum of integrals in $k$-space, but it can be simplified as $C_{2z}$ eigenvalues due to the existence of \G$_x$ and \G$_y$~\cite{Kim-nonprimitive,Kim-PC}, which will be explained in detail in the following subsection.

%%%% add PRR

\subsection{\ZZ{} topological invariant $\nu$ for the \G-protected TCI phase for space group \#110}
\label{sectionIVA}
Under the glide symmetry $\mathcal{G}_y=\{M_y|00\frac{1}{2}\}$, the \ZZ{} glide invariant $\nu$ characterizing the TCI phase protected by glide symmetry is written as 
a sum of integral terms within the $k$ space. It has been shown that $\nu$ can be in an expression of symmetry-based indicator by adding an additional inversion symmetry, i.e., by considering space groups \#13, \#14 and \#15, which is calculated only by the irreducible representations at high-symmetry points \cite{Kim2019prb100,Kim-nonprimitive,Kim-PC}. 

\begin{figure}[h]
\centering
\includegraphics[scale=0.36]{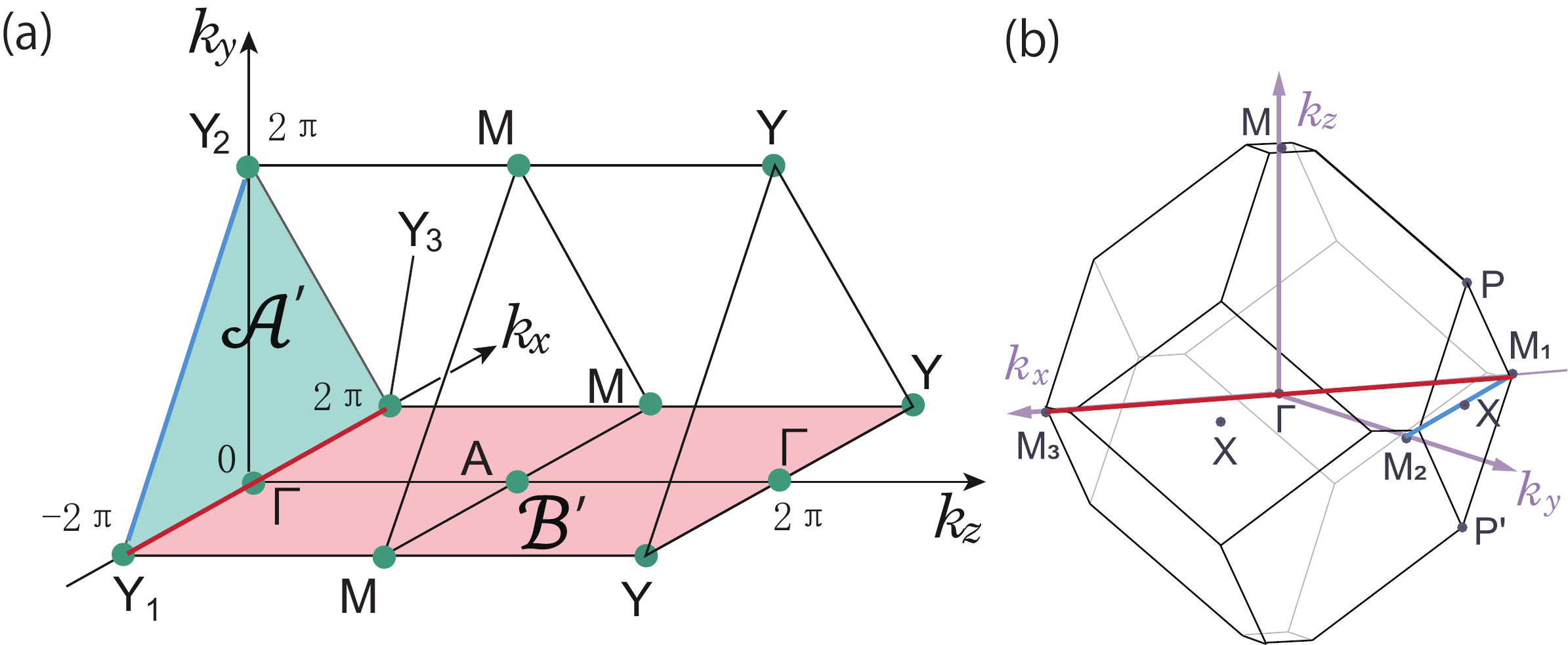}
\caption{Brillouin zones for \#9 and \#110 used in the calculation of the glide-\ZZ{} invariant $\nu_y$. (a) Half of the Brillouin zone of the space group \#9. 
(b) Brillouin zone of the space group \#110.
Blue and red lines are the paths for the Berry phases used in the calculation of $\nu_y$.}
\label{fig:bz}
\end{figure}

Here, we focus on spinless systems with space group \#110, which has two glide symmetries \G$_x=\{M_x|00\frac{1}{2}\}$ and \G$_y=\{M_y|00\frac{1}{2}\}$. In \#110, we can define two \G-protected topological invariants $\nu_x$ and $\nu_y$ associated with \G$_x$ and \G$_y$, 
respectively. 
%(We note the fractional translations are different defined in the main text for \LBO{} with \G$_y=\{M_y|\frac{1}{2}\frac{1}{2}0\}$ and  \G$_x=\{M_x|\frac{1}{2}\frac{1}{2}0\}$, but they become identical by a shift of the origin, and they share the same results). 
In the following, we will show that $\nu_x$ and $\nu_y$ are equal and given by $n/2$ mod 2, where $n$ is the number of
occupied bands. 
Since \#110 is a body-centered tetragonal lattice, the formula of $\nu$ here will be greatly different with that in primitive lattices. 
Thus, we will start with the formula of $\nu_y$ for the \G$_y$ on \#9 having the base-centered lattice
as follows \cite{Kim-PC}:
\begin{align}
\nu_y &= \frac{1}{2\pi} \left[ \int_{\mathcal{A}'} F_{xy} dk_xdk_y + \int_{\mathcal{B}'} \left( \frac{1}{2} F_{zx} - F_{zx}^+ \right) dk_zdk_x \right] \nonumber \\
& \quad + \frac{1}{\pi} \left( \gamma_{Y_1Y_2} - \gamma^+_{Y_1Y_3} \right) .
\label{eq:z2glide-9-2}
\end{align}
where $\mathcal{A}'=\{\bm{k}|k_z=0,\ 0<k_y,k_x+k_y<2\pi,k_y-k_x<2\pi\}$ and $\mathcal{B}'=\{\bm{k}|k_y=0,-2\pi<k_y<2\pi,0<k_z<2\pi\}$ are depicted in Fig.~\ref{fig:bz}(a) and $Y_1(-2\pi,0,0)$, $Y_2(0,2\pi,0)$, and $Y_3(2\pi,0,0)$. 
Here the formula is altered with a shift along $k_z$ direction by $\pi$ from that in Ref.~\cite{Kim-PC} for convenience.

To obtain the formula for \#110, we first make a coordinate transformation from the base-centered lattice to the body-centered tetragonal lattice. This is straightforward because \#9 is a subgroup of \#110
Then with the other glide symmetry \G$_x$, we show that three surface-integral terms vanish by symmetry constraints, with remaining term as
\begin{align}
&\nu_y = \frac{1}{\pi} \left( \gamma_{M_1M_2} - \gamma^+_{M_1M_3} \right) .
\end{align}
where $M_1(-2\pi,0,0)$, $M_2(0,2\pi,0)$, and $M_3(2\pi,0,0)$
are TRIM of \#110 (see Fig.~\ref{fig:bz}(b)).
In spinless systems, we can utilize the $C_{2z}$ symmetry as a result of \G$_{x}$ and \G$_y$ symmetries (Such a rule is broken in the spinful case, which is because the commutation relations between \G$_x$ and \G$_y$ changes. This is consistent with the ill-definedness of \Q{} in spinful systems with two glide symmetries). Since $C_{2z}$ and \G$_y$ commute with each other, we can evaluate $e^{i\pi\nu_y}$ via the sewing matrix below:
\begin{align}
(-1)^{\nu_y}=\prod_i \frac{\zeta_i(M)}{\zeta_i(X)} \frac{\zeta^+_i(\Gamma)}{\zeta^+_i(M)}.
\end{align}
Here $\zeta$ is the $C_{2z}$ eigenvalues and $\zeta^{+}$ is the $C_{2z}$ eigenvalues for the \G$_x=+e^{-ik_z/2}$ sector.
Because $M$ and $\Gamma$ are on the common $C_{2z}$ invariant line, they share the same $C_{2z}$ and \G$_x$ eigenvalue.
Thus, $\nu_y$ can be simplified as
\begin{align}
(-1)^{\nu_y}=\prod_i \frac{\zeta_i(\Gamma)}{\zeta_i(X)}.
\label{eq:nu110}
\end{align}

%%%(**)
We note that there are two inequivalent $X$ points for \#110, but we do not need to distinguish them in Eq.~(\ref{eq:nu110}) because they are on the same $C_{2z}$ axis due to the non-primitive nature of the lattice. 
Therefore, we get: 
\begin{align}
&\nu_x=\nu_y, 
\end{align}%, and simply write $\nu$ in Eq.~(\ref{eq:Qnew}). 
 and we can simply write $\nu=\nu_x=\nu_y$ in the following. 
In fact, this formula is equal to the symmetry-based indicator $\mu_2$ for \T-breaking spinless systems with \#27 ($Pcc$2) having two 
glide symmetries \G$_x$ and \G$_y$, given by Ref.~\cite{ono2018unified}
\begin{align}
&\mu_2=\frac{1}{4}\sum_{\mathbf{k}:\ \mathrm{TRIMs\ at}\ k_z=0}(n_{\mathbf{k}}^+-n_{\mathbf{k}}^-).
\end{align}
We can directly show $\mu_2=\nu$ as follows. 
In \#27, there are four TRIM on $k_z=0$: $\tilde{\Gamma}$ $(0,0,0)$,
$\tilde{M}$ $(\pi,\pi,0)$, $\tilde{X}$ $(\pi,0,0)$, $\tilde{Y}$ $(0,\pi,0)$. 
The Brillouin zone for \#27 is a half of that for \#110, and each of these four TRIM in \#27 
corresponds to two $k$ points in \#110, i.e., $\tilde{X}$ point (and also $\tilde{Y}$) correspond to two non-TRIM points in 
\#110. 
Such correspondence shows that two states with $C_{2z}=1$ and with $C_{2z}=-1$ present in \#27 will give no contribution to $\mu_2$. 
 The $\tilde{M}$ point corresponds to $X$ $(\pi,\pi,0)$ and $X^{\prime}$ $(-\pi,\pi,0)$ 
  in \#110, which are in fact on the common $C_{2z}$ axis, sharing the same $C_{2z}$ eigenvalue. 
Similarly, $\tilde{\Gamma}$ point corresponds to $\Gamma$ $(0,0,0)$ and $M$ $(2\pi,0,0)$ 
  in \#110, which are also on the common $C_{2z}$ axis, sharing the same $C_{2z}$ eigenvalue. Thus, we obtain
\begin{align}
\mu_2&=\frac{1}{4}\{2(n_{\Gamma}^+-n_{\Gamma}^-)+2(n_{X}^+-n_{X}^-)\}\nonumber\\
&\equiv -(n_{\Gamma}^-+n_{X}^-)\ (\mathrm{mod}\ 2),
\label{eq:mu2}
\end{align}
from which $\mu_2=\nu$ follows. 

At $\Gamma$, there are four 1D irreps $\Gamma_1,\cdots \Gamma_4$ with $C_{2z}=1$ and one 2D irrep
$\Gamma_5$ with two states having $C_{2z}=-1$. Thus, all the irreps contribute trivially to the product of $\nu$. 
On the other hand, the irrep at the $X$ point is 2D one with opposite $C_{2z}$ eigenvalues, therefore, the product for $\nu$ is equal to
\begin{align}
\nu=N/2\ \pmod{2}, 
\end{align}
 where $N$ is the number of occupied bands. 
Thus, $\nu$ is nontrivial when $N=4m+2$ ($m$: integer) and trivial when $N=4m$. 

In fact, it is also discussed in Ref.~\cite{ono2018unified} that $\nu$ is solely determined by the filling of the system in \#106 and \#110, which are supergroups of \#27. 
Topological nature of this topologically nontrivial phase has not been understood so far \cite{ono2018unified}. 
However, as shown in our paper, this symmetry-based indicator $\mu_2$ is in fact the \G-protected topological invariants $\nu_x$ and $\nu_y$. 
In this subsection, we have shown these properties for gapped systems. In the next subsection, we will see them also for gapless systems. 

\subsection{Topological invariant $\nu$ (=\Q{}) for the Dirac semimetal phase with two vertical \G{}}
\label{sectionIVB}

When time-reversal symmetry is preserved in space group \#110, the states are always fourfold degenerate at the momenta of \ZZ{} Dirac point. Thus, one cannot have a gap with $n=4m+2$ ($m$: integer). 
A \T-breaking perturbation is required for the topological crystalline insulator~(TCI) phase with nontrivial $\nu$, as we discussed in the previous section. 
Thus we consider the case where the \T{} is slightly broken to open a tiny bulk gap, where we can safely apply the bulk-surface correspondence for the \G-protected topological invariants $\nu_x$ and $\nu_y$.

%, and a curvature of the Fermi arc in $k$-space abruptly change as the energy changes. 
%table
\begin{table}[t] 
\begin{tabular}{lccc}
\hline
{\begin{tabular}[c]{@{}l@{}}Space\\ group\end{tabular}} & {\begin{tabular}[c]{@{}c@{}}Two vertical\\ glide mirrors\end{tabular}}         & {Location} & {\begin{tabular}[c]{@{}c@{}}Momenta used\\ for Eq.~(\ref{eq:nu110}) \end{tabular}} \\ \hline \hline
\#73                               & \{$M_x|\frac{1}{2},\frac{1}{2},0$\}; \{$M_y|\frac{1}{2},0,0$\}                & W    & $\Gamma$, T                                                                       \\
\#110                             & \{$M_x|\frac{1}{2},\frac{1}{2},0$\}; \{$M_y|\frac{1}{2},\frac{1}{2},0$\}  & P     & $\Gamma$, X                                                               \\
\#142                             & \{$M_x|\frac{1}{2},\frac{1}{2},0$\}; \{$M_y|\frac{1}{2},0,0$\}               & P  & $\Gamma$, X                                                                              \\
\#206                             & \{$M_x|\frac{1}{2},\frac{1}{2},0$\}; \{$M_y|\frac{1}{2},0,0$\}               & P    & $\Gamma$, N                                                                       \\
\#228                             & \{$M_x|\frac{1}{2},\frac{3}{4},0$\}; \{$M_y|\frac{3}{4},\frac{1}{2},0$\}   & W    & $\Gamma$, X                                                           \\    
\#230                             &  \{$M_x|\frac{1}{2},\frac{1}{2},0$\}; \{$M_y|\frac{1}{2},0,0$\}              & P  & $\Gamma$, N                                              \\                    
\hline  
\end{tabular} 
\caption{Spinless systems where \ZZ{} Dirac points associated with QHSSs. 
%Each column represents for the space group number, two vertical glide mirrors, the location for $Z_2$ Dirac points and the momenta used for Eq.~(\ref{eq:Qnew}) %\footnote{We note that all the space groups listed here have a wallpaper group of $p2gg$ on the (001) surface with two vertical glide mirror symmetries, and surface with $p4gg$ is forbidden to have QHSSs.}.
 }
\label{tab:1}    
\end{table}

In the present case of gapless systems, $\nu$ in Eq.~(\ref{eq:nugamma}) is ill-defined because the presence of the bulk Dirac point on the blue lines in 
Fig.~\ref{fig:ZZ} (b1), but instead, we can use Eq.~(\ref{eq:nu110}) to safely define $\nu$.
%Thus, we conclude bulk-surface correspondence for the topological invariant $\nu$ in Eq.~(\ref{eq:Qnew}) also holds. 
%leading to DHSSs. 
%This argument is based on the continuity of the surface states and of the topological invariant $\nu$ through a slight breaking of \T{}-symmetry.
Equation (\ref{eq:nu110}) is a new formula for \Q{}$(=\nu)$ calculated by the symmetry data at TRIM, and as we have seen that it is also equal to the symmetry-based indicator $\mu_2$ in Eq.~(\ref{eq:mu2}). 
$\mu_2$ is also entangled with filling-enforced topological crystalline insulators when there is full gap of the system, because $\mu_2$ is related to the filling $N$ by $\mu_2\equiv N/2 \pmod{2}$ in the space group \#110. Therefore, insulators with $N=4m+2\ (mod\ 2)$ ($m$: integers) are nontrivial in \#110 \cite{Kim-PC}.

We note Eq.~(\ref{eq:nu110}) still holds for other space groups listed in Tab.~\ref{tab:1} with proper high-symmetry momenta. 
Furthermore, only \#73, \#110 and \#142 can be converted to filling-enforced topological crystalline insulators when \T-symmetry is broken. 

\subsection{{BSC for \Q{} with two vertical \G{} in spinless systems}}
\label{sectionIVC}
\begin{center}
\begin{figure}
\includegraphics[scale=1.0]{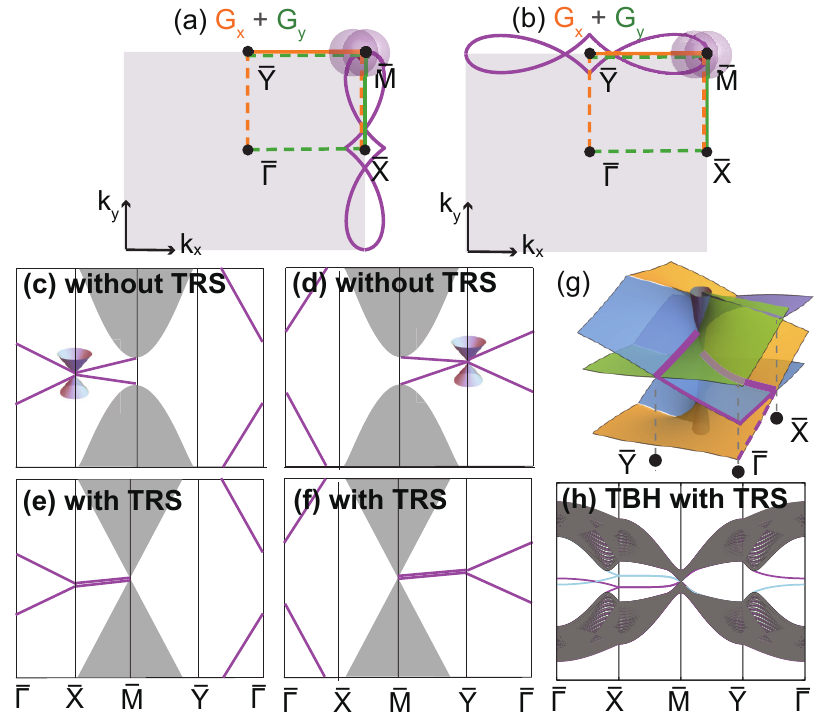}\caption{\ZZ{} Dirac points with \T- and two vertical \G-symmetries and associated QHSSs. 
(a) Locations and Fermi arcs for \ZZ{} Dirac points on the surface BZ preserving $\mathcal{G}_{x}$ and $\mathcal{G}_{y}$. Green (orange) dashed/solid lines are $\mathcal{G}_{y(x)}$-invariant/$\tilde{\Theta}_{y(x)}^2=-1$ lines. Purple solid lines are Fermi arcs, which will change to (b) when the energy changes.  
(c-d) Two possible surface-state connections for quad-helicoid surface states after breaking \T, which corresponds to topological crystalline insulators with two nontrivial \G-protected topological invariants $\nu_x=\nu_y=1$. Surface Dirac cones located at $\bar{X}$ in (c) and $\bar{Y}$ in (d) will evolve to QHSSs contributed by nontrivial \Q{} in (e) and (f), respectively. 
(e-f) Two different surface-state connections in \T-preserving systems with nontrivial \Q, which correspond to the Fermi arcs shown in (a) and (b), respectively. ``TRS'' represents for time-reversal symmetry. 
(g) Illustration for quad-helicoid surface states, where the purple lines show the gapless nature of the surface states along $\bar{Y}$-$\bar{\Gamma}$-$\bar{X}$ directions. 
(h) Surface states calculation obtained by the tight-binding model for \#110 with two vertical glide symmetries, where two \ZZ{} Dirac points are projected onto $\bar{M}$. Purple and blue lines represent the surface states from the top surface and bottom surface of the slab, which show quantum spin Hall like flows along $\bar{Y}$-$\bar{\Gamma}$-$\bar{X}$ directions, resulting in quad-helicoid surface states. 
}
\label{fig:gz2}
\end{figure}
\end{center}

%By combining our results on systems with one \G, we show bulk-surface correspondence for \T{}-preserving systems with two vertical \G. 
Now we will show BSC for \Q{} with two vertical \G{}, similarly to cases with one \G{} in Sec.~\ref{sectionIIID}.
Figures~\ref{fig:gz2} (a-b) show the surface BZ on (001) surface preserving two vertical \G.
In \T-preserving spinless systems with two vertical \G, \ZZ{} Dirac points are projected to $\bar{M}$ on the (001) surface BZ, which makes $\nu_{x}^{\mathrm{surface}}$ and $\nu_{y}^{\mathrm{surface}}$ ill-defined. 
By borrowing similar analysis on systems with one \G, firstly, we introduce a \T-breaking perturbation to open a small gap for \ZZ{} Dirac points to make $\nu_{x}^{\mathrm{surface}}$ and $\nu_{y}^{\mathrm{surface}}$ well-defined and be equal to $\nu_x(=\nu_y)$, which will give rise to two possible nontrivial surface-state connections with a single surface Dirac cone pinned at $\bar{X}$ and $\bar{Y}$, respectively, as summarized in Figs.~\ref{fig:gz2} (c-d). 
Next, when the \T{} symmetry is restored,
doubly degenerate surface states along the surface BZ boundaries protected by \Th$_x^2=-1$ = \Th$_y^2$ will appear. 
The corresponding surface-state connections for QHSSs along high-symmetry lines are shown in Figs.~\ref{fig:gz2} (e-f).  

Here, doubly degenerate surface states also start exactly from the Dirac point at $\bar{M}$ and connect either to 
$\bar{X}$ or to $\bar{Y}$ with the protection of $\tilde{\Theta}_y^2=-1$ or $\tilde{\Theta}_x^2=-1$ along the BZ boundaries. Away from the $\bar{M}$-$\bar{X}$ or $\bar{M}$-$\bar{Y}$ line, this degeneracy will be split 
toward the conduction and valence bands, showing the QHSSs with two quantum spin Hall~(QSH) like flows along $\bar{Y}$-$\bar{\Gamma}$-$\bar{X}$ directions, as marked by the purple lines in Fig.~\ref{fig:gz2} (g). 
%In addition, $\bar{M}$ is composed of two \ZZ{} Dirac points with two pairwise anticrossing helical surface states.
%only a twofold degenerate surface state can exist at $\bar{M}$. 
This conclusion is supported by our tight-binding model calculation for \#110 shown in Fig.~\ref{fig:gz2} (h), where the purple (blue) lines represent the surface states obtained from the top (bottom) of the slab
(See Sec.~{S1} for details~\cite{supp}).  
The surface-state filling at $\bar{X}$/$\bar{M}$ and $\bar{X}^{\prime}$/$\bar{M}^{\prime}$ will be changed by one when the surface states cross through the \ZZ{} Dirac points, showing the helicoid nature of the surface states.

In spinful systems with two vertical \G, \Q{} for the bulk Dirac point is ill-defined
because of the presence of other bulk degeneracies at TRIM on the \Th$^2=-1$ planes, in contrast to spinless systems. Therefore, the BSC will also be eliminated.

Thus we have shown that the \ZZ{} monopole charge \Q{} characterizing MHSSs only has a global nature in $k$-space, and it is not a local quantity. This automatically guarantees that the MHSSs will survive perturbations which split the Dirac point into a nodal ring or a pair of Weyl points, as seen in the following material example. 
\begin{center}
\begin{figure*}
\includegraphics[scale=0.4]{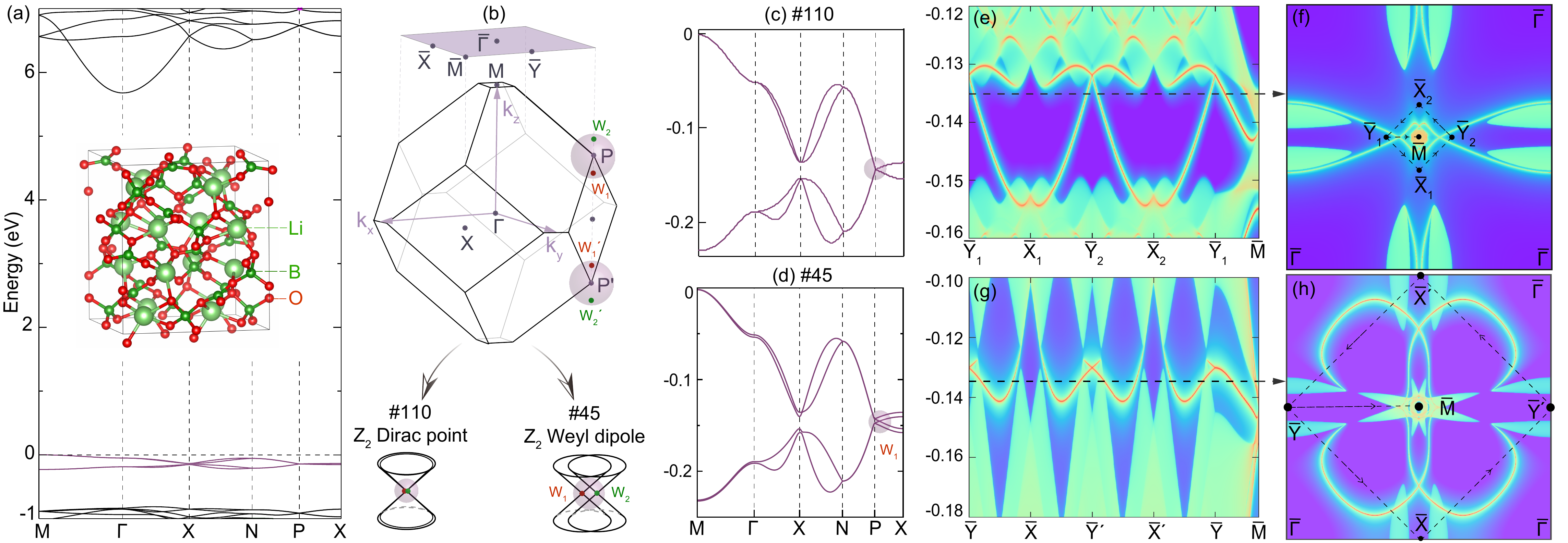}\caption{(a) Crystal structure and band structure for \LBO{}. (b) BZ and surface BZ for both \#110 and \#45 along [001] direction. (c-d) spinless band structure of \LBO{} for \#110 and \#45, respectively. (e-f) and (g-h) are QHSSs calculated on the (001) surface for \#110 and \#45, respectively, where the k paths in (e) and (g) are marked by the black dashed lines in (f) and (h), respectively. (f) and (h) are the Fermi arcs calculated around two \ZZ{} Dirac points located at $\bar{M}$. }
%(a) Crystal structure, BZ and (c) band structure for \LBO. (b) Surface states calculation along [001] direction, following the $k$ paths marked in (d). }
%BZ and surface BZ for both \#110 and \#45 along [001] direction. (c-d) spinless band structure of \LBO{} for \#110 and \#45, respectively. (e-f) and (g-h) are QHSSs calculated on the (001) surface for \#110 and \#45, respectively, where the $k$ paths are followed by the ones marked in (f) and (h). (f) and (h) are the Fermi arcs calculated around two \ZZ{} Dirac points located at $\bar{M}$, with the energy contour marked by the black dashed line in (e) and (g), respectively.}
\label{fig:LBO110}
\end{figure*}
\end{center}

\section{{\ZZ{} Dirac material \LBO{} with QHSSs}} 
\label{sectionV}

QHSSs are the consequence of BSC for \ZZ{} Dirac points with two vertical \G, as well as the bridge for connecting \Q{}, $\nu$ and $\mu_2$. %Thus, discovering materials with QHSSs will not only offer platforms to verify our theory, but also be essential for a wide range of further experimental and theoretical studies.
By following Tab.~\ref{tab:1}, we propose the first experimentally synthesized material candidate \LBO{} with \#110~\cite{singh2011synthesis} and two $Z_2$ Dirac fermions in its spinless electronic band structure, associated with QHSSs. 
As shown in Figs.~\ref{fig:LBO110} (a) and (c), all the fourfold band crossings at $P$ and $P^{\prime}$ are \ZZ{} Dirac points, which will be projected onto the corner point $\bar{M}$ on (001) surface. 
Figure~\ref{fig:LBO110} (b) is the surface state calculation on (001) surface, following the $k$-path marked in Fig.~\ref{fig:LBO110} (d). Two groups of anticrossing helical surface states with different Fermi velocity (chirality) together with degenerate surface states along surface BZ boundaries show the QHSSs feature of \LBO. 
%With breaking $C_{4z}^{\prime}$ symmetry, two \ZZ{} Dirac points split into two pairs of \ZZ{} Weyl dipoles along $k_{z}$ direction. %Moreover, the (001) surface preserves the plane group of $pgg$, surface states will be doubly degenerate along all the surface BZ boundary lines, forming an inner nodal chain in the reciprocal space, protected by $\tilde{\Theta}_x^{2}=\tilde{\Theta}_y^{2}=-1$. 

QHSSs can be also obtained in systems with \ZZ{} Weyl dipoles. 
After breaking $C_{4z}^{\prime}$ symmetry, 
%(such as shifting the oxygen atoms along $z$ direction or simply add a strain along $x$ direction)
the symmetry of \LBO{} is lowered to \#45 and two \ZZ{} Dirac points split into two pairs of \ZZ{} Weyl dipoles along $k_{z}$ direction, i.e., $W_{1}+W_{2}$ and $W_{1}^{\prime}+W_{2}^{\prime}$, respectively, as shown in Figs.~\ref{fig:LBO110} (b) and (d). 
Since $\tilde{\Theta}_x$ and $\tilde{\Theta}_y$ are still preserved in \#45, each pair of Weyl dipole will also carry a nonzero \Q{} and leads to QHSSs on the (001) surface, as shown in Figs.~\ref{fig:LBO110} (g) and (h), which are quite similar with the ones from two \ZZ{} Dirac points.  
Within a pair of the Weyl dipole, $W_{1}$ and $W_{2}$ cannot be annihilated with each other due to the nonzero \Q{}, although their total monopole charge $\mathcal{C}$ is zero, which is different from general Weyl semimetals. 

%Furthermore, Fermi arcs connect Weyl points only from two $\mathcal{T}$-related Weyl dipoles on the surface, which is different from normal Weyl semimetals.

%

\section{Conclusion}
{We offer a full understanding on the topology of \GT-protected \ZZ{} Dirac points in this paper. We offer the gauge-invariant formula for \ZZ{} monopole charge \Q{} protected by (\GT)$^2=-1$, and find \Q{} can be only defined globally in the $k$-space. 
\Q{} can be formulated into a simpler form in terms of irreducible representations at two high-symmetry momenta when two vertical \G{} are present. 
DHSSs can be obtained in both spinless and spinful systems with one \G{}, while QHSSs can be only obtained in spinless systems with two vertical \G{}. 
QHSSs in spinless \ZZ{} Dirac systems is also the bridge to unify \GT-protected \Q{}, \G-protected $\nu$, \G-protected symmetry-based indicator $\mu_2$ and even filling-enforced topological crystalline insulators. We also offer a material candidate \LBO{} and a list of space groups having QHSSs.}

\section*{{Acknowledgements}}
We acknowledge the supports from Tokodai Institute for Element Strategy (TIES) funded by MEXT Elements Strategy Initiative to Form Core Research Center Grants No. JPMXP0112101001, JP18J23289, JP18H03678, and JP22H00108. T. Z. also acknowledge the support by Japan Society for the Promotion of Science (JSPS), KAKENHI Grant No. JP21K13865.

\bibliographystyle{unsrt}
\bibliography{reference}
\newpage{}

\end{document}